\documentclass[referee]{raa}           
\usepackage{graphicx,times}
\usepackage{natbib}
\usepackage{cancel}
\usepackage{amssymb,amsmath}
\usepackage{soul}
\usepackage{xcolor}
\bibpunct{(}{)}{;}{a}{}{,}

\usepackage[a4paper=true, pagebackref=true]{hyperref}
\hypersetup{pdftitle = The title of my PDF, pdfauthor = My name, pdfsubject= The subject, pdfkeywords = keyword1 keyword2 keyword3} 
\hypersetup{colorlinks = true, linkcolor = black, anchorcolor = red, citecolor = blue, filecolor = red, pagecolor = red, urlcolor = red}

\begin{document}

   \title{The first symbiotic stars from the LAMOST survey
}

 \volnopage{ {\bf 2015} Vol.\ {\bf X} No. {\bf XX}, 000--000}
   \setcounter{page}{1}

   \author{Jiao Li\inst{1, 2}, Joanna Miko{\l}ajewska\inst{3}, Xue-Fei Chen\inst{1}, A-Li Luo\inst{4}, Alberto Rebassa-Mansergas\inst{5}, Yonghui Hou\inst{6}, Yuefei Wang\inst{6}, Yue Wu\inst{4}, Ming Yang\inst{4}, Yong Zhang\inst{6} and Zhan-Wen Han\inst{1}
   }

   \institute{Key Laboratory for the Structure and Evolution of Celestial Objects, Yunnan observatories, Chinese Academy of Sciences, P.O. Box 110, Kunming, 650011, China; {\it lijiao@ynao.ac.cn, zhanwenhan@ynao.ac.cn}\\
        \and
             Chinese Academy of Sciences University, China\\
         \and
             Nicolaus Copernicus Astronomical Center, Bartycka 18, PL 00–716 Warsaw, Poland\\
        \and
             National Astronomical observatories, Chinese Academy of Sciences, Beijing 100012, China\\
        \and
             Kavli Institute for Astronomy and Astrophysics, Peking University, Beijing 100871, China\\
         \and
             Nanjing Institute of Astronomical Optics \& Technology, National Astronomical Observatories, Chinese Academy of Sciences, Nanjing 210042, China\\
\vs \no
  {\small Received --; accepted --}
}

\abstract{Symbiotic stars are interacting binary systems with the longest orbital periods. They are typically formed by a white dwarf, a red giant and a nebula. These objects are natural astrophysical laboratories for studying the evolution of binaries. Current estimates of the population of Milky Way symbiotic stars vary from 3000 up to 400000. However, the current census is less than 300. The Large sky Area Multi-Object fiber Spectroscopic Telescope (LAMOST) survey can obtain hundreds of thousands of stellar spectra per year, providing a good opportunity to search for new symbiotic stars.  In this work we detect 4 of such binaries among 4,147,802 spectra released by the LAMOST, of which two are new identifications. The first is LAMOST J12280490-014825.7, considered to be an S-type halo symbiotic star. The second is LAMOST J202629.80+423652.0, a D-type symbiotic star.
   }

   \authorrunning{J. Li et al. }            
   \titlerunning{Symbiotic stars observed by LAMOST}  
   \maketitle

%
\section{Introduction}           
\label{sect:intro}

Symbiotic stars are interacting binary systems with the longest orbital periods and a multicomponent structure, that is an evolved cool giant and an accreting hot, luminous companion (usually a white dwarf, WD; some might be a neutron star or a black hole) surrounded by a dense ionized symbiotic nebula. Therefore, symbiotic stars have composite spectra. 

Depending on the nature of the cool giant, symbiotic stars are generally classified into three types: S-type (stellar), D-type (dusty) and D$'$-type.  S-types contain normal giant stars of effective temperatures $T_{\rm eff} \sim3000-4000$~K. They are the most numerous ($\sim 80\%$) and their orbital periods are of the order of a few years.  D-types contain Mira-type variables and warm ($T\sim1000$~K) dust shells. Their orbital periods are larger than 15 years (\citealt{Joanna2003}). D$'$-types contain F- or G-type cool giant stars surrounded by a dust shell with a temperature not higher than $500$~K (\citealt{Allen1984}). There are only 8 D$'$-types out of 188 symbiotic stars in the \cite{Belczynsk2000} catalogue. The presence of both an evolved giant that suffers from heavy mass loss and a hot component copious in ionizing photons results in rich and luminous circumstellar environment. In particular, very different surroundings are expected, including both neutral and ionized regions, dust forming regions, accretion disks, interacting winds and jets, which makes symbiotic stars a natural laboratory for studying the binary star interaction and evolution (\citealt{Corradi2003}, and references therein).

Symbiotic stars are proposed as probable progenitors of type Ia supernovae (SNe Ia) (\citealt{Hachisu1999}; \citealt{Chen2011}), as rapid mass accretion from the cool giant component may lead the WD to reach the Chandrasekhar mass limit (\citealt{Han2004};  \citealt{Wangbo2012}). SNe Ia have been successfully used as a cosmological distance indicator (\citealt{Riess1998}; \citealt{Perlmutter1999}), however, the exact nature of the progenitors of SNe Ia is still not clear. SNe Ia PIF 11kx, with a broad H$\alpha$ line in its late spectrum, may result from a symbiotic star (\citealt{Dilday2012}). Large numbers of symbiotic stars can help us to study the progenitors of SNe Ia. Hence, searching for new symbiotic stars is of extreme importance.

Current estimates for the total Galactic symbiotic population vary considerably from 3000 (\citealt{Allen1984}), 30000 (\citealt{Kenyon1993}) to 3-4$\times 10^5$ (\citealt{Magrini2003}).  However, there are less than 300 Galactic symbiotic stars known (\citealt{Belczynsk2000}; \citealt{Miszalski2013}; \citealt{Miszalski2014}; \citealt{Rodriguez2014}). This incompleteness is due to the lack of deep all-sky surveys at wavelengths where symbiotic stars can be efficiently selected against other stellar sources (\citealt{Belczynsk2000}; \citealt{Miszalski2014}; \citealt{Joanna2014}). Most of the known symbiotic stars have been discovered by large H$\alpha$ surveys, e.g. the INT Photometric H$\alpha$ Survey and the AAO/UKST SuperCOSMOS H$\alpha$ Survey, and then confirmed by the deep spectroscopy and long-term I-band light curves of the candidates. The Large sky Area Multi-Object fiber Spectroscopic Telescope (LAMOST) survey can obtain hundreds of thousands of stellar spectra per year, providing a good opportunity to find new symbiotic stars. Since no effort has been made from the large spectroscopic observation survey, we devise a selection method to find out symbiotic stars from the data released by LAMOST (including DR1, DR2 and DR3). As a result, we detect 4 symbiotic star spectra, of which two are new identifications, the other two have been previously identified by \cite{Downes1988}.

The structure of the paper is as follows. Section~\ref{sect:Obs} briefly introduces the LAMOST survey and our method to select symbiotic stars. Section~\ref{sect:4_syst} describes the 4 symbiotic stars we identify. Section~\ref{discussion} is the discussion. We conclude in Section~\ref{sect:conclusion}.

\section{Spectroscopic observations and target selection}
\label{sect:Obs}

\subsection{The LAMOST Survey}
\label{LAMOST}

LAMOST is a quasi-meridian reflecting Schmidt 4.0 m telescope located in Xinglong Station of National Astronomical Observatories, China, characterized by both large field of view and large aperture. The telescope is equipped with 16 low-resolution spectrographs, 32 CCDs and 4,000 fibers. Each spectrograph is fed with the light from 250 fibers and approximately covering a wavelength range $3700-9100 $~\AA\ at a resolving power of about 1800. With the capability of its large field of view and  a large number of targets, LAMOST telescope dedicates for spectral survey of celestial objects in the whole northern sky (\citealt{Cui2012}; \citealt{Zhao2012}).

The current data of LAMOST contain data release 1 (DR1), data release 2 (DR2) and data release 3 (DR3). The number of DR1, DR2 and DR3 spectra are 2204860, 1443843 and 499099, respectively (see Table~\ref{tab1}). The spectra are classified as stars, galaxies, quasars and unknowns (mainly due to the poor quality of their spectra). DR1 includes both 1,487,200 spectra obtained during the first year of the general survey, and 717,660 spectra which were observed during the pilot survey (\citealt{Luo2012}). Due to the lack of network of photometric standard stars, the flux calibration is relative. The LAMOST spectra are unambiguously identified by a plate identifier, a spectrograph identifier and a fiber identifier (\citealt{Song2012}; \citealt{Ren2014}).

Based on the LAMOST data, lots of works have been done. For example, \cite{Ren2014} identified 119 white dwarf-main sequence binaries, 
\cite{Zheng2014} reported a hypervelocity star with a heliocentric radial velocity of about $620$~km/s, and
\cite{Zhong2014} discovered 28 candidate high-velocity stars at heliocentric distances of less than $3$~kpc. For more details see the website http://www.lamost.org/public/publication?locale=en.

 \begin{table}
\bc
\begin{minipage}[]{100mm}
\caption[]{Data of LAMOST.\label{tab1}}\end{minipage}
\setlength{\tabcolsep}{7pt}
\small
 \begin{tabular}{llllll}
  \hline\noalign{\smallskip}
  \hline
Data& Total& Star& Galaxy& QSO& Unknown\\
  \hline\noalign{\smallskip}
DR1& 2,204,860& 1,944,406& 12,082& 5,017&243,355\\
DR2& 1,443,843& 1,240,792& 23,967& 4,251&174,833\\
DR3& 499,099& 437,299& 5,229& 574& 55,997\\
  \noalign{\smallskip}\hline
  \hline
\end{tabular}
\ec
\tablecomments{0.86\textwidth}{The number of spectra, which are
  classified as stars, galaxies, QSOs and unknowns by the LAMOST
  pipeline. DR1 includes both   1,487,200 spectra obtained during the first year of the general survey, and 717,660 spectra which were observed during the pilot survey. For more details see the website http://dr.lamost.org/ucenter/mytoken.}
\end{table}

\subsection{Optical spectra of symbiotic stars}
\label{Spectra}

Optical spectra of known symbiotic stars have been catalogued by \cite{Allen1984} and \cite{Munari2002}. These spectra are characterized by the presence of absorption features typical of late-type giant stars (spectral type K or M). These include, among others, CaI, FeI, H$_{\rm 2}$O, CO and TiO lines. Strong nebular Balmer line emission, as well as HeI, HeII and the [OIII], [NeIII], [NV] and [FeVII] forbidden lines can also be seen.  Additionally, more than $50\%$ of symbiotic stars show broad emission features at $\lambda6830$~\AA\, and a factor of $\sim 4$ weaker but similar emission feature at $\lambda7088$~\AA, which only exists in symbiotic stars.  The $\lambda6830$~\AA\, band profile has a typical width of about $20$~\AA\, and it often ranks among the 10 most intense lines in the optical region of symbiotic stars (which may reach $5\%$ of the intensity of H$\alpha$).  The $\lambda\lambda6830$ and $\lambda\lambda7088$ lines are due to Ramma scattering of the OVI resonance doublet $\lambda\lambda1032$, $\lambda\lambda1038$ by the neutral hydrogen in the atmosphere of the cool giant and the inner parts of its stellar wind (\citealt{Schmid1989}).

\begin{figure}
   \centering
   \includegraphics[width=14.0cm, angle=0]{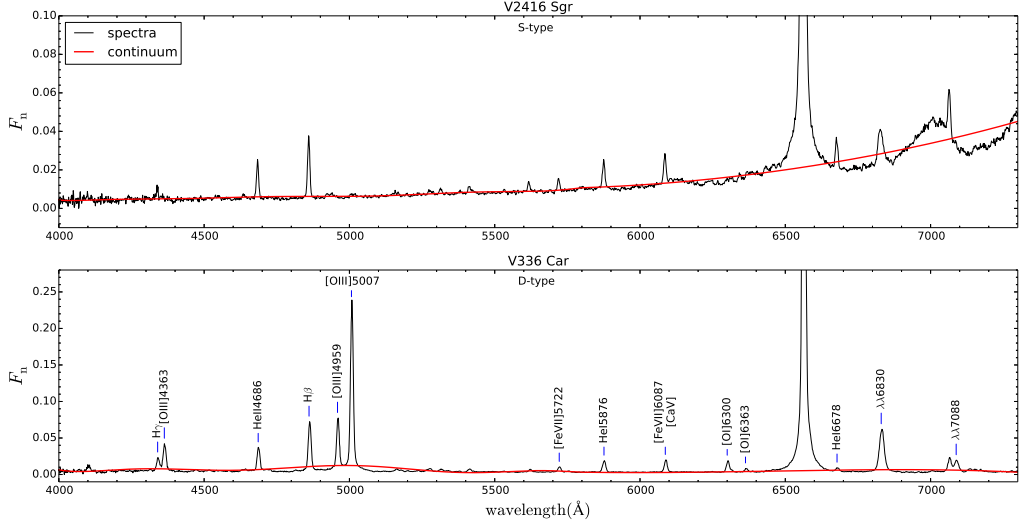}
   \caption{V2416 Sgr (top panel) and V336 Car (bottom panel),
     an S-type and a D-type symbiotic star, respectively (\citealt{Allen1984b}). The black
     and red lines are the spectra (continuum plus lines)
     and continua of the objects, respectively. We have normalized the
     flux into the range of [0, 1] following
     Eq.~(\ref{normalize}). }
   \label{Fig1}
   \end{figure}

In S-type stars (e.g. V2416 Sgr in Figure~\ref{Fig1}, top panel), the continuum has obvious absorption features originating from the red giant. Whilst the Balmer, HeI and HeII lines are strong, the forbidden lines such as [OIII], [NeIII] are weak and even fade away. The Balmer decrement is generally steeper than ${\rm H}\alpha/{\rm H}\beta \sim2.8$ (case B) (\citealt{Emission-Line_Stars}).

In D-type stars (e.g. V336 Car in Figure~\ref{Fig1}, bottom panel), the Balmer lines are strong and the forbidden lines are well developed in a wide range of excitation from [OI] up to [FeVII] and [NeV]. The Balmer decrement is also steeper than in the nebular case. The wide range of ionization levels in D-types reflects the fact that their nebulae are ionization-bounded whereas in at least some S-types the nebulae are density-bounded.

D$'$-type stars (see Figure~\ref{StHA190}; note that this is one of our 4 identified LAMOST symbiotic binaries) have similar emission lines as for D-types. The most conspicuous property of D$'$-type stars is the presence of a G- or F-type giant instead of a cooler star (\citealt{Murset1999}).

According to above described features, we use the following criteria to classify an object as a symbiotic star (\citealt{Joanna1997}; \citealt{Joanna2014}; \citealt{Belczynsk2000}):

{\bf (a)} Presence of the absorption features of a late-type giant:
TiO, H$_{\rm 2}$O, CO, CN and VO absorption bands, as well as CaI, CaII, FeI and NaI lines.

{\bf (b)} Presence of strong HI and HeI emission lines and
presence of emission lines from ions such as HeII, [OIII], [NV] and [FeVII] with an equivalent width exceeding $1$~\AA.

{\bf (c)} Presence of the $\lambda\lambda6830$ emission line, even if the cool giant features are not obvious.

\subsection{Selection Method}
\label{method}

Based on the spectral properties of symbiotic stars and the feature of LAMOST data, we devise a selection method to find symbiotic stars. This follows a three-step procedure.

{\bf (i)} In order to directly compare the spectra, we first normalize the flux into the range of [0,1] following
\begin{equation}
\label{normalize}
 F_{\rm n}
=\frac{F_{\lambda} - F_{\rm min}}{F_{\rm max} - F_{\rm min}},
\end{equation}
where $F_{\rm max}$, $F_{\rm min}$ are the maximum and minimum fluxes of each spectrum, respectively.

{\bf (ii)} Unlike the vast majority of spectra observed by LAMOST, symbiotic stars display Balmer emission lines, especially the H$\alpha$ and the H$\beta$ lines. We therefore select symbiotic stars by analyzing the intensity of the H$\alpha$ and H$\beta$ lines, i.e. $I({\rm H}\alpha)$ and $I({\rm H}\beta)$. These intensities are obtained from the following equation,

\begin{equation}
\label{normalize}
 I(\lambda_0)
=\int _{ \lambda_0 -\Delta\lambda  }^{ \lambda_0 +\Delta\lambda  }{ \left( F_{\rm n}-F_{\rm c} \right)d\lambda  } 
\end{equation}

where $F_{\rm c}$ is the continuum fitted by the Global Continuum Fit method of \cite{Leesunyoung2008} (the [OIII]$\lambda4340, \lambda4959$ and $\lambda5007$ lines are removed when we fit the continuum). The $\Delta\lambda$ for H$\alpha$, H$\beta$ are set to $50$ and $10$~\AA, respectively. As the fluxes have been normalized, the intensity of the lines do not have truly physical significance.

\begin{figure}
   \centering
   \includegraphics[width=14.0cm, angle=0]{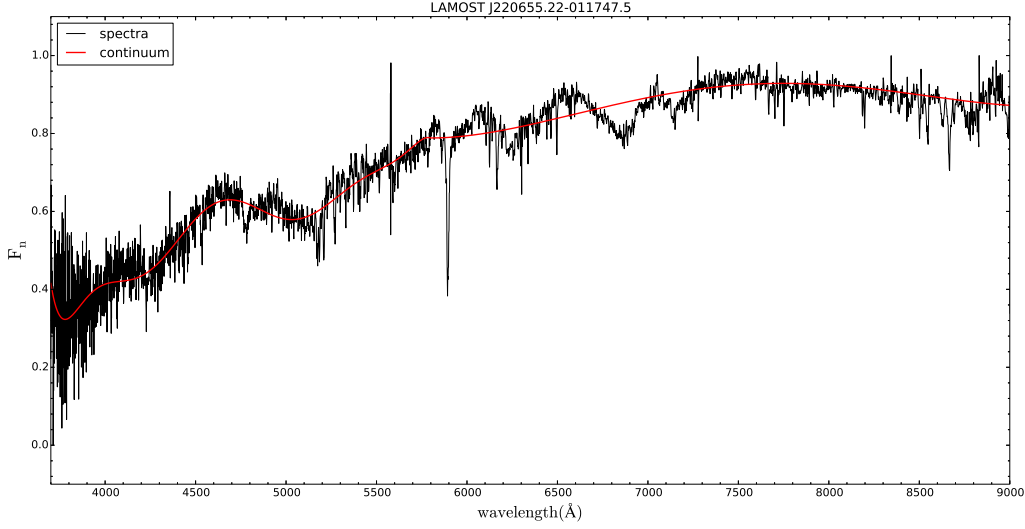}
   \caption{LAMOST J220655.22-011747.5, is an M0 type spectrum.
   The black line stands for the spectra of the object. The red line
   is the continuum fitted by the Global Continuum Fit method of \cite{Leesunyoung2008}.}
   \label{M_fit}
   \end{figure}

It is important to emphasize that fitting the continuum of M-type or later type spectra is subject to systematic uncertainties at the $\lambda5500-7500$~\AA\, range, as shown in Figure~\ref{M_fit}. In order to efficiently select symbiotic stars, it becomes important to set proper values for the intensity of the H$\alpha$ and H$\beta$ lines. In our work, they were set to $I({\rm H}\alpha)>0.5$ and $I({\rm H}\beta)>0.2$, which efficiently exclude spectra displaying the Balmer series in absorption or in faint emission (such as the spectrum shown in Figure 2).

{\bf (iii)} The Balmer series and other emission lines in symbiotic stars are very prominent, hence the continuum is generally flat in the optical spectrum (see Figure~\ref{Fig1}). We therefore average the normalized continuum $F_{\rm n}$ in the $4000-7300$~\AA\, wavelength range and exclude all spectra with the average value being above 0.4.

After performing the above steps over the entire LAMOST spectroscopic database, we select 4758, 7631 and 2377 spectra in DR1, DR2, and DR3, respectively, as symbiotic star candidates (see Table~\ref{tab1_1}). Visual inspection of the spectra revealed that most of these were planetary nebula and B[e] stars. Finally, we find four symbiotic stars, two of which have been catalogued in the past, the other two are new identifications (see Table~\ref{tab2}).

\begin{table}
\bc
\begin{minipage}[]{110mm}
\caption[]{Spectra selected by our routine as symbiotic
  binaries\label{tab1_1}}\end{minipage} \setlength{\tabcolsep}{7pt}
\small
 \begin{tabular}{llllllc}
  \hline\noalign{\smallskip}
  \hline
Data& Total& Star& M-type&Step(ii)& Step(iii)&Symbiotic star\\
  \hline\noalign{\smallskip}
DR1&2,204,860&1,944,406&67,282&44,983&4,758&1\\
DR2&1,443,843&1,240,792&54,888&64,691&7,631&2\\
DR3&499,099&437,299&18,054&15,408&2,377&1\\
  \noalign{\smallskip}\hline
  \hline
\end{tabular}
\ec
\tablecomments{0.86\textwidth}{(1) Total stands for the number of the
  spectra we have used in each data release of the LAMOST. (2)
  Star, M-type columns stand for stars and M type spectra,
  respectively. (3) Step(ii) and Step(iii) columns stand for the
  selected out spectra by step(ii), step(iii) in
  Section~\ref{method}.}
\end{table}

\begin{table}
\bc
\begin{minipage}[]{100mm}
\caption[]{Symbiotic stars observed by LAMOST\label{tab2}}\end{minipage}
\setlength{\tabcolsep}{2pt}
\small
 \begin{tabular}{ccccccccccclc}
  \hline\noalign{\smallskip}
  \hline
Desig& Date& IR type& $u$ &$g$ &$r$ &$i$ & $z$&Cool spectra& DR& Name\\
 (LAMOST J)&MJD&&&&&&&&&\\
  \hline\noalign{\smallskip}
122804.90-014825.7&56592&S&15.88&15.46&12.18&11.67&11.92&K3&DR2&---\\
194957.58+461520.5&56592&S&12.54&---&---&---&---&M2&DR2&StHA 169\\
202629.80+423652.0&56946&D&15.05&14.07&13.49&---&---&Mira&DR3&---\\
214144.88+024354.4&56204&D$'$&10.94&10.33&10.01&11.39&10.56&G5&DR1&StHA 190\\
  \noalign{\smallskip}\hline
  \hline
\end{tabular}
\ec
\tablecomments{0.86\textwidth}{Basic properties of the symbiotic
  stars, which are ordered by right ascension (R. A.) of the epoch
  J2000.0. The first and third symbiotic stars have been indentified
  by \cite{Downes1988}, the others are new symbiotic stars.}
\end{table}

\section{Symbiotic stars observed by LAMOST}
\label{sect:4_syst}

Figures 3--6 show the LAMOST spectra of the 4 symbiotic
stars we have identified.
 
\subsection{LAMOST J122804.90-014825.7}
\label{122804.90-014825.7}
   
\begin{figure}
   \centering
   \includegraphics[width=14.0cm, angle=0]{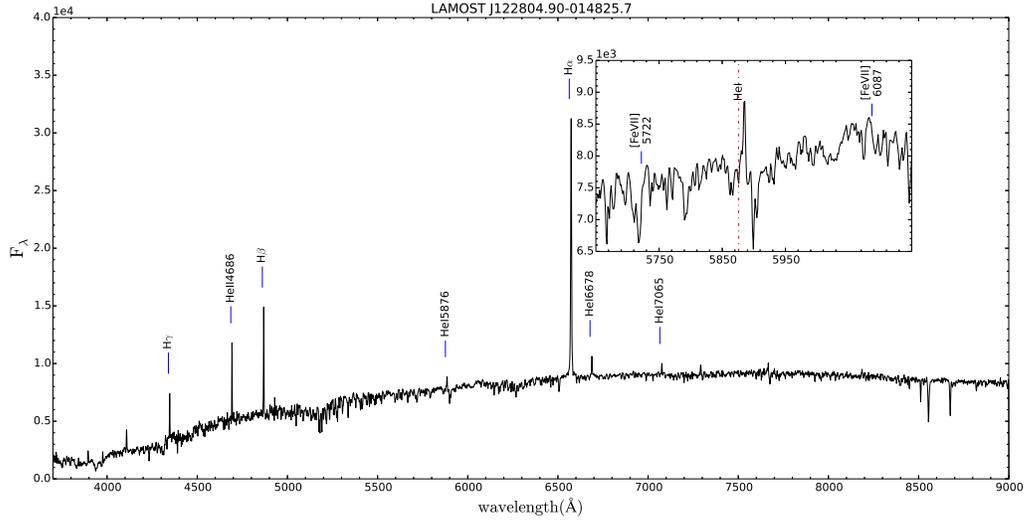}
   \caption{LAMOST J122804.90-014825.7 is a new S-type symbiotic star,
     with a radial velocity $v_{\rm
       r}\sim\left(374\pm2\right)$~km/s. The vertical dashed line in the
     inset stands for the wavelength of HeI $\lambda5876$. The faint
     [FeVII] lines have the same velocity shift. $F_\lambda$
   is relative flux.}
   \label{LAMOSTJ12280490-0148257}
   \end{figure}

The presence of strong HeII $\lambda4686$ emission implies the presence of a hot component in the object (Figure~\ref{LAMOSTJ12280490-0148257}). The spectrum is with a radial velocity of $v_{\rm r}\sim\left(374\pm2\right)$~km/s. And we can see the faint [FeVII] emission lines with the same Doppler shift in the inset of Figure~\ref{LAMOSTJ12280490-0148257}. The spectrum displays also Mg$\lambda5176$ and CaII triplet absorption lines. It is classified as K3 type by the LAMSOT pipeline, which stands for the cool giant component of a symbiotic star. So we propose it is an S-type system.

Based on the $g$, $r$ and $i$ magnitudes, the V magnitude is $m_{\rm V}\sim13.5$~mag (http://classic.sdss.org/dr4/algorithms/sdssUBVRITransform.html), and we assume a typical absolute magnitude of the K3 type cool giant to be $M_{\rm V}\sim0.1$~mag. The spectroscopic distance in this case would be about $5$~kpc. The Galactic coordinates are $l=291.0062$, $b=60.5226$, so the vertical distance between the object and the Galactic disk is about $4.3$~kpc. It seems to be a halo symbiotic star.

\subsection{LAMOST J194957.58+461520.5}
\label{StHA169}

The LAMOST spectrum of J194957.58+461520.5 is very similar to the one presented by \cite{Downes1988} for the $4000-7300$~\AA\, range. LAMOST J194957.58+461520.5 is included in the \cite{Belczynsk2000} catalogue of symbiotic stars as StHA 169. The spectrum shows a strong and broad H$\alpha$ emission line. The other Balmer lines, as well as HeI and HeII$\lambda4686$ lines are also clearly present. At the same time there is a faint feature at $\lambda\lambda6830$ as shown in Figure~\ref{StHA169}. The spectrum displays obvious TiO absorption bands and CaII triplet absorption lines, and the star is classified as M2 by the LAMOST pipeline. It is an S-type symbiotic star.

\begin{figure}
   \centering
   \includegraphics[width=14.0cm, angle=0]{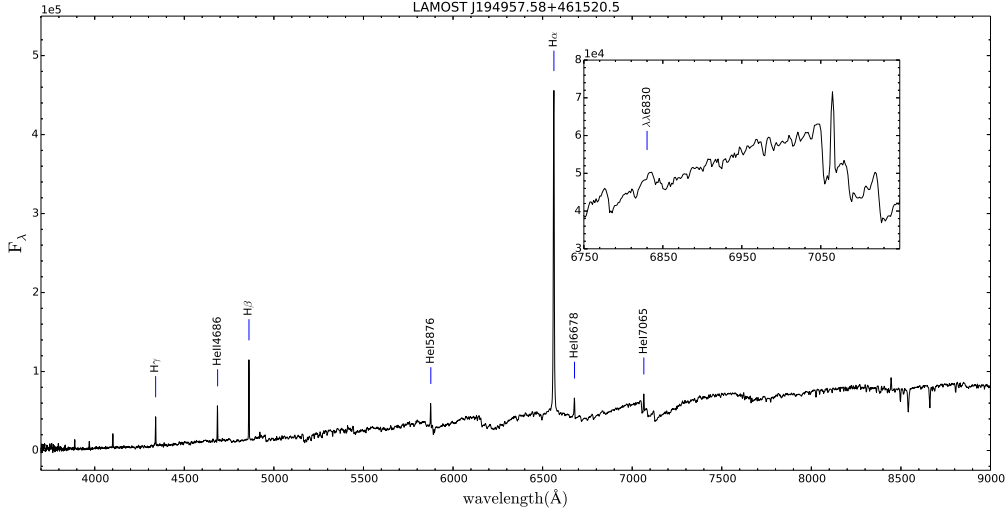}
   \caption{LAMOST J1194957.58+461520.5 is a S-type symbiotic star,
     which has been observed by \cite{Downes1988}, and listed as StHA
     169 in \cite{Belczynsk2000} catalogue of symbiotic stars. $F_\lambda$
   is relative flux.}
   \label{StHA169}
   \end{figure}

\subsection{LAMOST J20262980+4236520}
\label{20262980+4236520}

\begin{figure}
   \centering
   \includegraphics[width=14.0cm, angle=0]{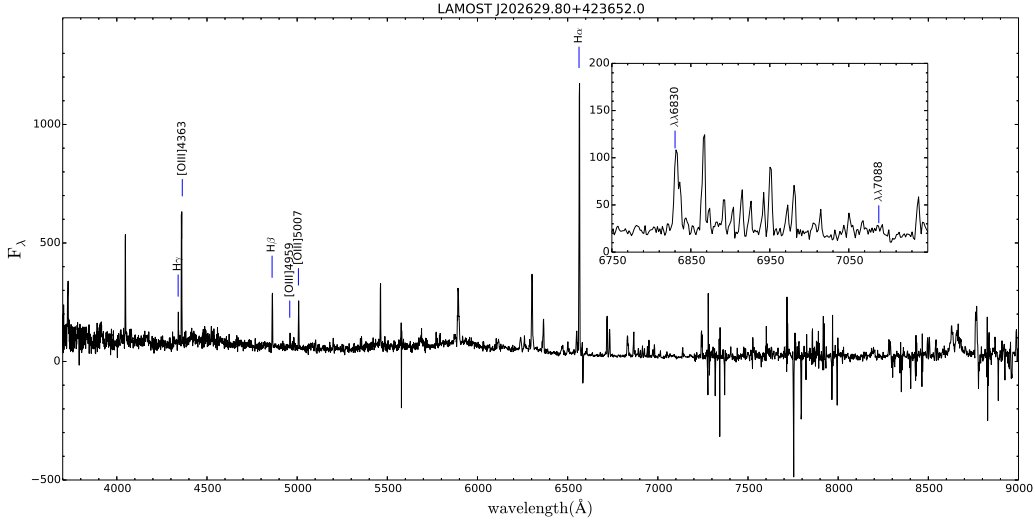}
   \caption{LAMOST J20262980+4236520 is a new D-type symbiotic
     star. The red part of the spectrum is affected by systematic uncertainties due to sky subtraction. $F_\lambda$ is relative flux.}
   \label{LAMOSTJ20262980+4236520}
   \end{figure}

The spectrum of this object is shown in Figure~\ref{LAMOSTJ20262980+4236520}. The Balmer and [OIII] lines are clearly present. Weak [NII] $\lambda6548$, $\lambda6584$ lines at the both sides of the H$\alpha$ line can also be seen, however, the HeI and HeII lines are very weak. There is an obvious $\lambda\lambda6830$ line ($I\left(\lambda\lambda6830\right)/I\left({\rm H}\alpha\right)\sim7.6\%$) and a $\lambda\lambda7088$ line. The flat continuum, and the intensity ratio $I\left({\rm[OIII]}\lambda5007\right)/I\left({\rm H}\beta\right)\sim0.89$ and $I\left({\rm[OIII]}\lambda4363\right)/I\left({\rm H}\gamma\right)\sim4.29$ (\citealt{Gutierrez1995}; \citealt{Joanna2014}; \citealt{Rodriguez2014}) suggests that LAMOST J20262980+4236520 is a D-type symbiotic star, so its cool giant is a variable. The signal to noise ratio (S/N) is relatively low at the $iz$ bands where the spectrum is also affected by systematic uncertainties due to sky subtraction.

\subsection{LAMOST J214144.88+024354.4}
\label{214144.88+024354.4}

\begin{figure}
   \centering \includegraphics[width=14.0cm,
     angle=0]{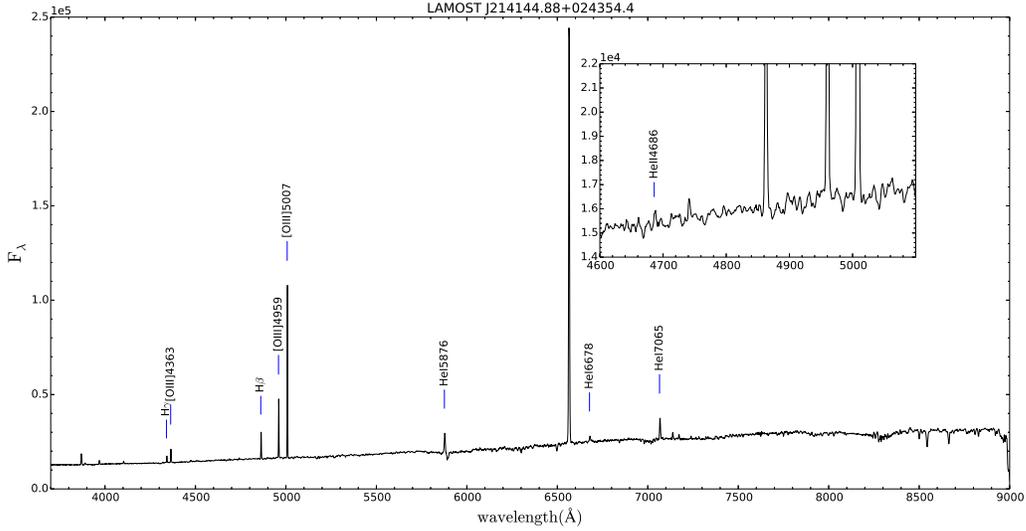}
   \caption{LAMOST J214144.88+024354.4 is a D$'$-type symbiotic star, included in the \cite{Belczynsk2000} list as StHA 190. $F_\lambda$ is relative flux.}
   \label{StHA190}
   \end{figure}

LAMOST J214144.88+024354.4 is a D$'$-type symbiotic star. It is included in the catalogue by \cite{Belczynsk2000} as StHA190. The spectrum, shown in Figure~\ref{StHA190}, is nearly identical to the one presented by \cite{Downes1988} in the $4000-7300$~\AA\, wavelength range. The spectrum is classified as G5 type by the LAMOST pipeline, which is consistent with the D$'$-type nature of this symbiotic binary. High-resolution optical spectra of StHA 190 were analyzed by \cite{Smith2001}, who derived the giant temperature, $T_{\rm eff}=5300\pm150$~K and ${\rm [Fe/H]}\sim0$ based on fitting the spectrum with model atmospheres. The spectral type reported here is consistent with these results. 

\section{Discussion}
\label{discussion}
We used 4,147,802 LAMOST spectra to find symbiotic stars, of which 3,622,479 are classified as stars by the LAMOST pipeline (see Table~\ref{tab1_1}). Since about $1\%$ of the stars are red giants in our solar neighborhood (http://pages.uoregon.edu/jimbrau/astr122/Notes/Chapter17.html), a rough estimate of the number of the red giant stars in these spectra is around 30,000. \cite{Magrini2003} proposed that the number of symbiotic stars can be taken to be $0.5\%$ of the total number of red giants. Therefore there should be more than 100 symbiotic stars to be selected out. But we should note that the estimated number of Galactic symbiotic stars from \cite{Magrini2003}  is 1-2 orders of magnitude higher than that given by others as shown in Section~\ref{sect:intro}. So the real symbiotic stars among the LAMOST stellar spectra may be less than 100. As the first step, we only identified 4 symbiotic binaries, far below what we expected. By comparing the coordinates between the known symbiotic stars listed in the catalogue of \cite{Belczynsk2000} and the footprint of LAMOST DR1, there should be 9 known symbiotic stars in the area of the footprint (see Table~\ref{syst_footprint} ). According to the plate design of the LAMOST, for brighter stars ($r \le 14$~mag), only those with $9 \le J \le 12.5$~ mag from the 2MASS catalogs are chosen as potential targets (Luo et al. 2015). It immediately rules out 8 of the 9 symbiotic stars. For the fainter stars ($r > 14$~mag), there are 5 conditions to select potential targets (\citealt{Yuan2015}), which might exclude the symbiotic star Draco C-1. The StHA 190 has not been excluded, probably because its J-band magnitude is very close to the target selection condition of brighter stars. In the future work, we will apply the selection method to the SDSS (the Sloan Digital Sky Survey) released data to find more symbiotic stars and check its efficiency.

\begin{table}
\bc
\begin{minipage}[]{100mm}
\caption[]{Symbiotic stars in the footprint of LAMOST DR1\label{syst_footprint}}\end{minipage}
\setlength{\tabcolsep}{6pt}
\small
 \begin{tabular}{cccccc}
  \hline\noalign{\smallskip}
  \hline
Name& RA& Dec & $J$ &$r$\\
  \hline\noalign{\smallskip}
EG And&00 44 37.1&+40 40 45.7&3.65&6.6\\
BD Cam&03 42 09.3&+63 13 00.15&1.31&4.5\\
UV Aur&05 21 48.8&+32 30 43.1&4.03&8.3\\
TX CVn&12 44 42.0&+36 45 50.6&7.47&8.7\\
T CrB&15 59 30.1&+25 55 12.6&5.70&2.1\\
AG Dra&16 01 40.5&+66 48 09.5&7.15&9.2\\
Draco C-1&17 19 57.6&+57 50 04.9&14.38&16.7\\
StHA 190&21 41 44.8&+02 43 54.4&8.71&10.0\\
AG Peg&21 51 01.9&+12 37 29.4&5.00&5.8\\
  \noalign{\smallskip}\hline
  \hline
\end{tabular}
\ec
\tablecomments{0.86\textwidth}{The symbiotic stars are in the area of the footprint of LAMOST DR1. RA and Dec are the celestial coordinates.}
\end{table}

\section{Conclusion}
\label{sect:conclusion}
We have found 4 symbiotic stars among 4,147,802 spectra released by the LAMOST DR1, DR2 and DR3. The objects LAMOST J194957.58+461520.5 and LAMOST J214144.88+024354.4 are included in the \cite{Belczynsk2000} catalogue of the symbiotic stars as StHA 169 and StHA 190, respectively, and they can be selected out by our method. The LAMOST spectra of these two objects are nearly identical to those provided by \cite{Downes1988}.  LAMOST J12280490-014825.7 is a new symbiotic star and is considered to be a Halo S-type with a radial velocity of $v_{\rm  r}\sim\left(374\pm2\right)$~km/s. LAMOST J20262980+4236520 is also a new symbiotic star classified as D-type. With the operating of the LAMOST, more and more symbiotic star spectra will be obtained.

\normalem
\begin{acknowledgements}
This work was supported in part by the National Natural Science
Foundation of China (Grant Nos. 11390374, 11422324, 11173055, 11350110496), the Science and Technology Innovation Talent Programme of Yunnan Province (Grant No. 2013HA005), the
Talent Project of Young Researchers of Yunnan Province
(2012HB037), and the Chinese Academy of Sciences (Grant No. XDB09010202). ARM acknowledges financial support from the Postdoctoral Science Foundation of China (Grant Nos. 2013M530470, 2014T70010). The Guo ShouJing Telescope (the Large Sky Area 
Multi-Object Fiber Spectroscopic Telescope, LAMOST) is a National
Major Scientific Project built by the Chinese Academy of
Sciences. Funding for the project has been provided by the National
Development and Reform Commission. LAMOST is operated and managed by
the National Astronomical Observatories, Chinese Academy of
Sciences. The LAMOST Data release Web site is http://data.lamost.org/.

\end{acknowledgements}
  
\bibliographystyle{raa}
\bibliography{syst_in_lamost}

\end{document}